# Event-Aware Loss Design for Forecasting of Convective Precipitation and Lightning

ChangJae Lee[a,b], Heecheol Yang[c], Byeonggwon Kim[a]

[a] *Forecast Bureau, Korea Meteorological Administration, Seoul, Korea*

[b] *M.S. in Data Science Program, The University of Texas at Austin, Austin, TX, USA*

[c] *Distributed Intelligence and Systems Laboratory, Division of Computer Convergence, Chungnam National University, Daejeon, Korea.*

*Corresponding author*: Heecheol Yang, hcyang@cnu.ac.kr

ABSTRACT

Accurate forecasting of high-impact weather, specifically extreme precipitation and lightning, remains a significant challenge in numerical weather prediction (NWP) due to the complexity of atmospheric microphysics. While deep-learning models have shown promise in large-scale forecasting, they often suffer from systematic under-prediction of rare, high-intensity events and localized convective showers when optimized with conventional loss functions like Mean Squared Error (MSE).

This study proposes an Event-Aware multi-task deep-learning post-processing framework designed to improve the representation of convective processes by leveraging lightning observations. The model jointly predicts precipitation amount, rainfall probability, and lightning occurrence using a shared-backbone Patch-cGAN (Conditional Generative Adversarial Network) architecture. To address the rare event problem, we introduce a lightning-informed loss-weighting strategy that element-wisely multiplies the MSE component by a spatial weight map derived from observed lightning strikes, forcing the model to prioritize accuracy in convective regions during training.

Evaluations conducted over the Korean Peninsula during the 2025 Summer demonstrate that our framework outperforms standard AI benchmarks and conventional NWP models, particularly at intense rainfall thresholds (≥ 40 mm/6 h). Furthermore, the model exhibits superior skill in predicting lightning compared to conventional lightning parameterization and instability-index-based methods. These results indicate that integrating physical event indicators into the loss formulation effectively guides models to learn the meteorological signatures of deep convection, offering a pathway toward more reliable extreme weather forecasting.

SIGNIFICANCE STATEMENT

Improving the accuracy of intense rainfall and lightning forecasts is vital for effective disaster response. While modern artificial intelligence (AI) has advanced weather prediction, standard models often fail to capture the extreme events and localized convective storms. This study introduces an Event-Aware deep-learning model that uses lightning observations to teach the AI where to prioritize its accuracy. By integrating these physical indicators of convective activity directly into the learning process, our model improves the forecasting of extreme convective rainfall and lightning. By incorporating physical observations into data-driven modeling, this research offers a more reliable method for predicting hazardous weather.

# 1. Introduction

*a. Motivation*

Accurate forecasts of precipitation and lightning are critical for mitigating flood risk, optimizing water-resource management, and supporting disaster preparedness and response. Nevertheless, they remain among the most challenging components of Numerical Weather Prediction (NWP). A primary source of this difficulty lies in the intrinsic complexity of atmospheric microphysical processes, which are typically approximated through parameterization schemes (Stensrud, 2007). These approximations introduce substantial uncertainty, particularly for convective precipitation systems.

In recent years, deep-learning approaches have emerged as an alternative paradigm for precipitation forecasting, enabling data-driven learning directly from observational and reanalysis datasets (e.g., Harris et al., 2022; Lam et al., 2023; Lang et al., 2024; Zhou et al., 2022). While these models have demonstrated skill in reproducing large-scale precipitation characteristics (Rasp et al., 2020; Rasp et al., 2023), they exhibit notable limitations, such as under-prediction of extreme events and convective, shower-type precipitation. These issues often stem from the nature of conventional loss functions, such as Mean Squared Error (MSE) or Generative Adversarial Network (GAN; Goodfellow et al., 2020) loss, which frequently struggle with rare event problems (Lee et al., 2025; Rasp et al., 2020; Rasp et al., 2023). Such limitations motivate the design of specialized loss functions regularizing the under-generation issues associated with the rare tail of the distribution.

In this study, we focus on lightning observations because most extreme precipitation and shower events are associated with convective activity. Lightning serves as a physically meaningful indicator of deep convection and intense updrafts (Deierling et al., 2008). This close relationship motivates the use of lightning not only as a primary forecast target but also as an event-aware supervisory signal during model training.

Accordingly, we propose a multi-task deep-learning–based post-processing framework that jointly predicts precipitation amount, rainfall probability, and lightning occurrence. In addition, we introduce a lightning-informed loss-weighting strategy, termed the Event-Aware Generative Adversarial Network (EA-GAN), that emphasizes meteorological conditions associated with convective events. By integrating lightning as both a predictive target and an event-aware learning signal, the proposed framework aims to improve the representation of convective precipitation processes, addressing the inherent limitations of conventional loss formulations.

This study demonstrates the effectiveness of the proposed event-aware loss by showing that it improves both the detection of convective shower events and the accuracy of extreme rainfall intensity predictions.

*b. Related Work*

A growing body of literature has explored deep-learning approaches for diagnosing or predicting precipitation directly from observational or reanalysis data. At the global scale, models such as GraphCast (Lam et al., 2023) and AIFS (Lang et al., 2024) have demonstrated the feasibility of end-to-end AI-based forecasting using ECMWF ERA5 reanalysis fields (Hersbach et al., 2020) at resolutions of approximately 0.25° × 0.25°. These models have shown strong performance for large-scale circulation and synoptic precipitation patterns.

At regional scales, higher-resolution precipitation estimates derived from satellite and radar observations have been widely used to train deep-learning models, enabling finer representation of mesoscale features (e.g., Harris et al., 2022; Zhou et al., 2022). Generative frameworks, including GAN and Diffusion-based models (Ho et al., 2020), have also been applied to improve the realism of precipitation fields by better matching observed intensity distributions (Harris et al., 2022; Lee et al., 2025; Vicens-Miquel et al, 2025).

Machine-Learning-based lightning forecasts models have also been explored (McClung et al., 2026). These approaches are specifically designed to provide accurate predictions of life-threatening and high-impact weather phenomena.

Despite these advances, difficulties in capturing convective, shower-type precipitation persist, particularly when models are trained on high-resolution observational datasets such as radar precipitation (Lee et al., 2025). This struggle may stem from its localized feature, which often yields contradicting results (i.e., rain or no rain) under the similar meteorological conditions. Case-based evaluations indicate that localized convective events are often under-detected or spatially smoothed, highlighting a fundamental challenge that extends beyond data resolution or model architecture alone. Furthermore, it has been widely reported that AI-based models utilizing an MSE loss function underestimate extreme events (Rasp et al., 2023).

Recent studies have introduced physically constraints, such as non-negative constraint for precipitation variable (Moldovan et al., 2025), and intensity-weighted loss functions (Vicens-Miquel et al, 2025) into the machine learning model, which leads to improvements in forecast skill. These approaches seek to improve forecast quality by adding domain-specific physical characteristics into existing architectures, rather than solely on increasing model parameters or expanding raining datasets.

To address the challenges associated with convective events more directly, the present study leverages lightning information to explicitly guide the learning process through an event-aware loss formulation. This approach improves both the detection of localized convective events and the prediction of extreme rainfall, which is strongly associated with intense convective activity. The effectiveness of this strategy is demonstrated through benchmark evaluations and detailed case studies presented in the subsequent sections.

## 2. Data and Methods

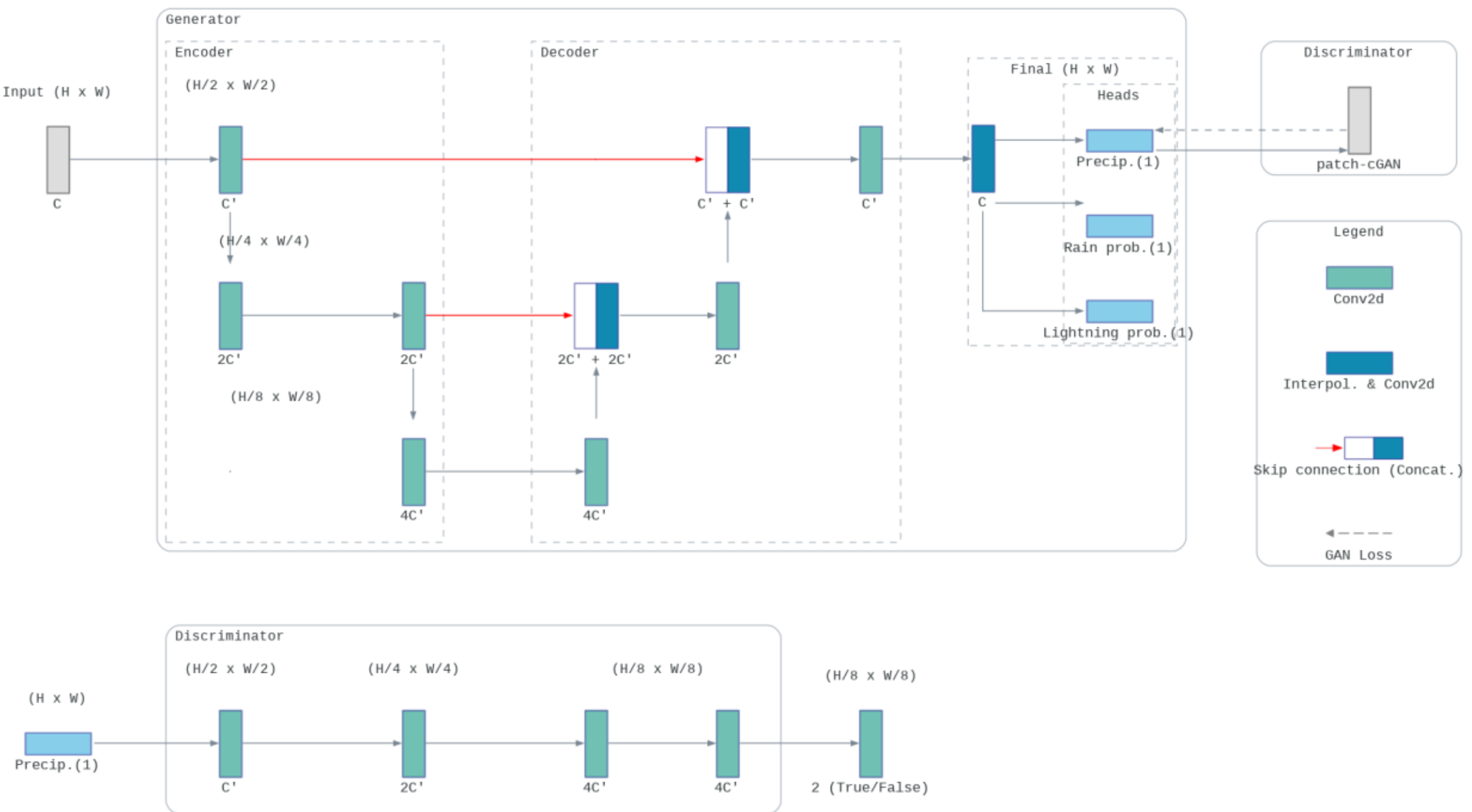

*Figure 1. Architecture of the proposed model. Each box represents a feature map, with the operations, such as convolution or skip connection, defined in the legend. The input data has dimensions of H (height) × W (width) × C (channels) (720 × 576 × 81), comprising meteorological variables from the NWP. The latent dimension C' is set to 128 within this architecture.*

In this study, we implement a multi-task deep-learning–based post-processing framework for precipitation and lightning forecasting. Schematic illustration is present in Figure 1. This study assumes a perfect NWP forecast, following the Perfect Prognostic Method (PPM; Marzban et al., 2006), whereby each observed precipitation and lightning are paired with the corresponding NWP input. By utilizing the PPM framework, the model characterizes the physical relationship between the meteorological predictors and the forecasting targets (rainfall and lightning). Consequently, this learned relationship can be extended to operational post-processing during inference, assuming the NWP inputs provide a sufficiently accurate representation of the atmospheric state at longer forecast lead times.

Dataset for training and validation is described in Table 1. Further details of the dataset are present in the coming sections.

| | Usage | Training period | Validation period |
|---|---|---|---|

| Numerical Weather Prediction | Input | 2021 - 2024 | 2025 |
|---|---|---|---|
| Radar estimated precipitation | Label | 2021 - 2024 | 2025 |
| Lightning observation | Label / Loss weight | 2021 - 2024 | 2025 |

*Table 1. Dataset description for the training and validation.*

*a. Numerical Weather Prediction: Input Data*

To implement a post-processing model, we take ECMWF's Integrated Forecasting System High RESolution forecast (IFS-HRES; Rasp et al., 2020), a global forecast product with 0.1° × 0.1° latitude/longitude resolution, as input data. Atmospheric variables of 15 pressure levels and the surface level, along with the altitude data obtained from the Shuttle Radar Topography Mission (SRTM; Farr et al., 2007), described in Table 2, are used here. For training, we use recent forecast data from 6h to 15h forecast lead times, considering a spin-up period to ensure stabilized NWP results (Ma et al., 2021), from the 00 and 12 UTC runs.

| Surface variables | Atmospheric variables | Pressure Level (hPa): 15 levels |
|---|---|---|
| 2m temperature<br>2m dew-point temperature<br>Mean sea level pressure<br>Surface net solar radiation<br>Altitude | U-component wind<br>V-component wind<br>Temperature<br>Relative humidity<br>Geopotential height | 1000, 950, 925, 900, 850, 800, 700, 600, 500, 400, 300, 250, 200, 150, 100 |

*Table 2. Input variables and levels for the proposed model.*

*b. Radar estimated precipitation: Data for target value (labeling)*

For the observational data, we use radar-estimated precipitation data processed with the CLEANER algorithm (Oh et al., 2020) from the Korea Meteorological Administration (KMA). This dataset has a spatial resolution of 500 m × 500 m and a temporal frequency of 10 minutes. 3-h accumulated precipitation is used as labeled data in this experiment.

In this experiment, 3-h accumulated precipitation labels are paired with single-step input data; for example, the 6-h forecast fields are utilized to predict the accumulated precipitation occurring between lead times of 3 and 6 hours. Alternatively, two successive time-steps could be employed as input to provide the model, to provide more temporally continuous predictions.

Figure 3 illustrates that radar-estimated precipitation closely aligns with rain gauge observations, which directly measure rainfall at the surface, compared to the reanalysis precipitation field (ERA5). The inconsistency between reanalysis fields and in-situ observations may arise from the low resolution of reanalysis data and its inherent limitations in convective parameterization. Figure 2 shows the locations of each radar site and its coverage, as well as the locations of the rain gauges. Since our proposed model uses radar-estimated precipitation as labels, we can expect the results to better capture the true nature of precipitation compared to models trained on reanalysis fields.

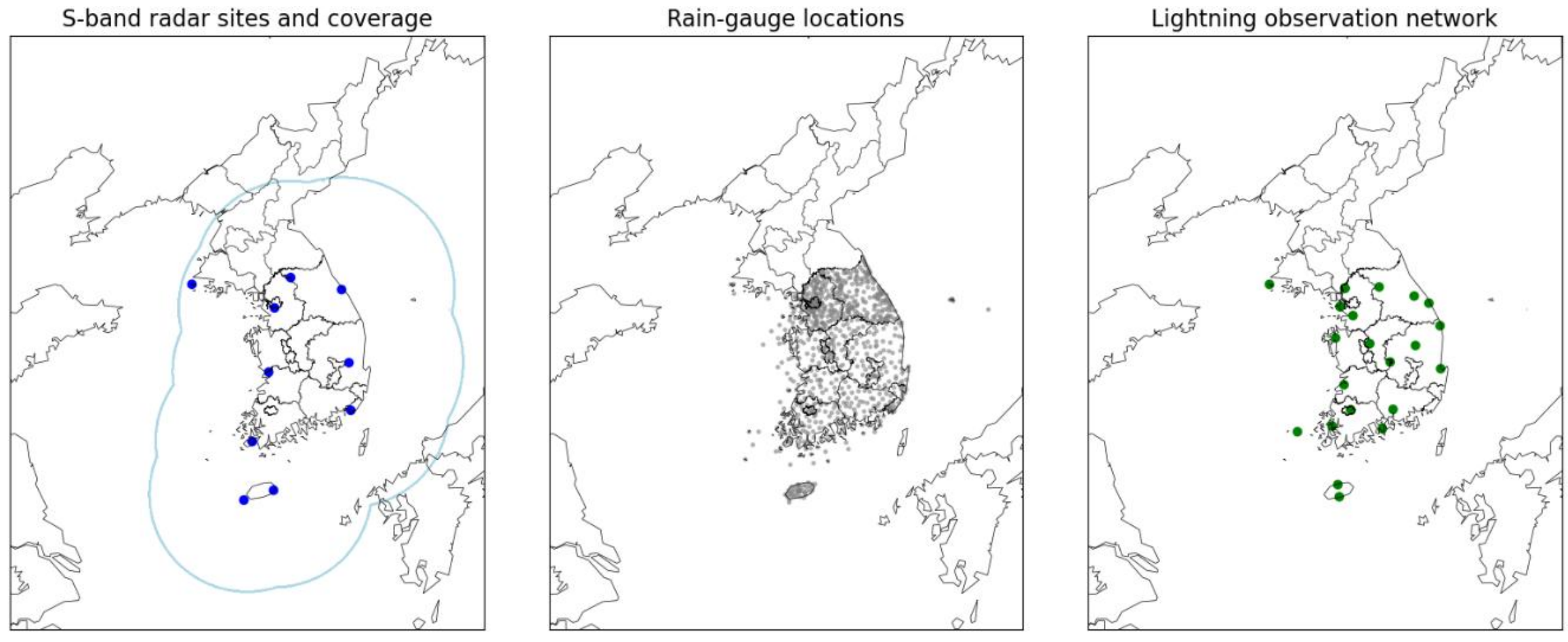

*Figure 2. (Left) Locations of S-band radar sites (blue dots) and their coverage areas (skyblue outline), (Middle) locations (gray dots) of rain gauges, and (Right) lightning observation network (green dots) operated by the Korea Meteorological Administration (KMA).*

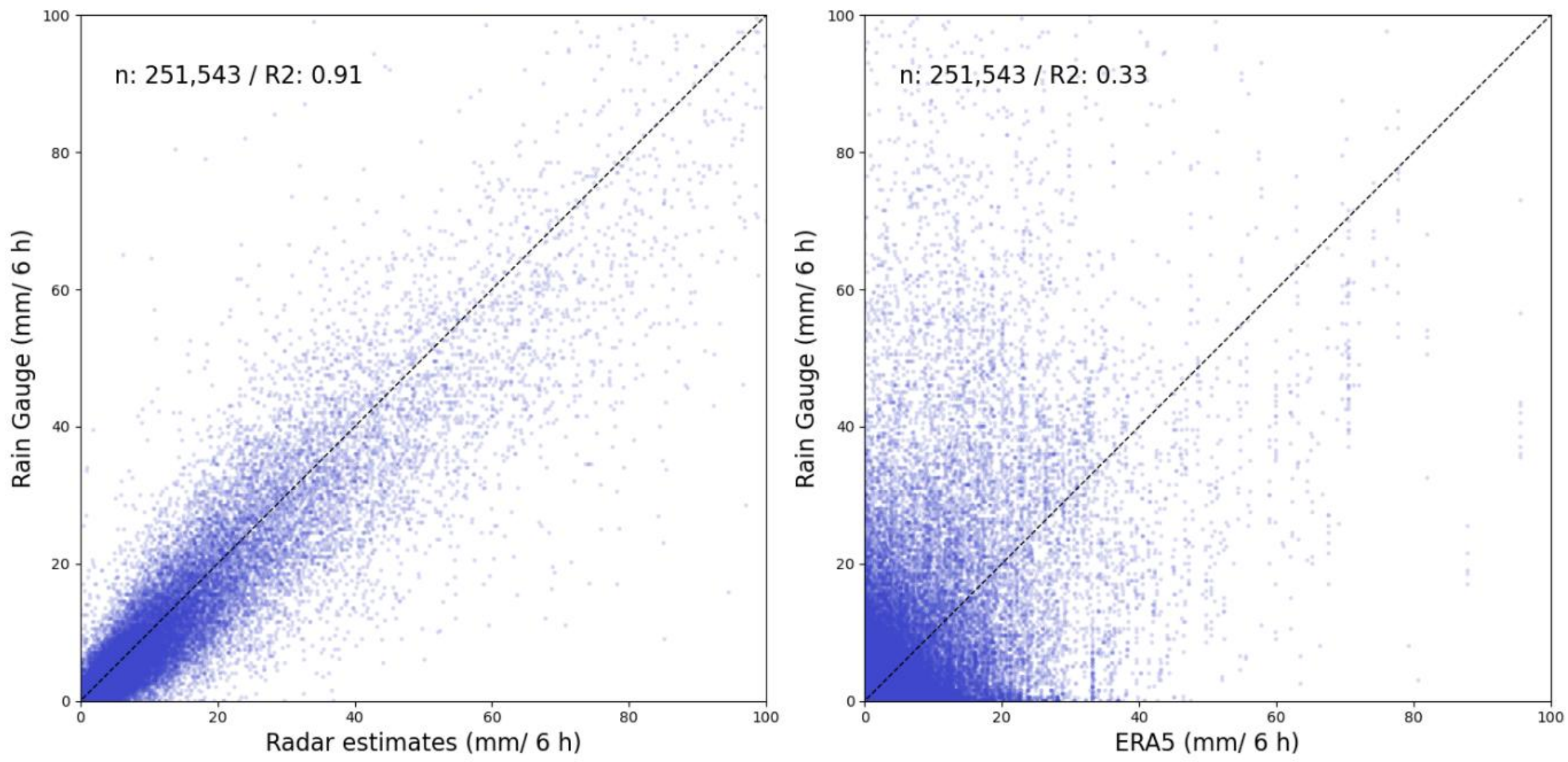

*Figure 3. (Left) Scatter plot comparing radar-estimated precipitation with observed rainfall from rain gauges, and (Right) comparison between ERA5-predicted precipitation fields and observed rainfall. Six-hour accumulated rainfall data from June to August 2024 is used. The comparison is based on the nearest-point method.*

*c. Lightning observation: Data for target (labeling) and loss weight*

To incorporate lightning information into the proposed framework, we utilize observational data obtained from a ground-based LIghtning NETwork (LINET; Betz et al., 2009). The network consists of 21 stations, as shown in Figure 2.

The detected lightning events are converted into a gridded representation consistent with the model output resolution. For each detected lightning location, a 10 km × 10 km grid cell centered at the strike position is masked as a lightning-occurrence region. This spatial buffering accounts for geolocation uncertainty and ensures consistency with the model grid spacing.

The resulting gridded lightning dataset serves a dual role within the framework; (1) the supervisory label for the lightning prediction task in the multi-task architecture and (2) the lightning-informed loss-weighting map that emphasizes convective regimes during training.

*d. Pre-processing input and labeled data*

We design our post-processing model to generate high-resolution precipitation fields conditioned on the input fields. However, due to the difference in spatial resolution and domain between the input (0.1° × 0.1°) and labeled data (2 km × 2 km), we re-grid them to a 2 km × 2 km resolution using the bilinear interpolation method and restricted the domain to around the Korean Peninsula region (1440 km × 1152 km).

After the re-gridding step, we normalize and standardize our input and labeled data to neural networks to be trained efficiently. The labeled data is log-transformed (Equation 1) before being normalized, since precipitation data has a highly-skewed distribution.

$$y' = (\log_{10}(y + c) - \mu_y)/\sigma_y, \quad (1)$$

where y is the raw precipitation amount (mm), c is a small constant (0.1) used to ensure numerical stability for zero-precipitation, $\mu_y$ and $\sigma_y$ are the mean and standard deviation, respectively, of the log-transformed precipitation values calculated from the training dataset, and $y'$ represents the resulting normalized precipitation label.

*e. Methods: Architecture*

The proposed model, as a multi-task NWP post-processing model, takes NWP fields as input and outputs precipitation amount, rainfall probability, and lightning occurrence. The architecture is based on Patch-cGAN (Conditional Generative Adversarial Network; Isola et al., 2017), comprising a U-Net-based generator (Ronneberger et al., 2015) and a convolutional discriminator. The Patch-cGAN architecture is selected to reconstruct a target distribution that closely aligns with the characteristics of the training dataset.

The generator adopts a U-Net encoder–decoder structure, and encodes meteorological variables as input and decodes them to produce corresponding fields for multi-tasks prediction. Its U-Net architecture captures both local and global features of input data through several consecutive convolutional layers with different sizes of kernels and residual skip connections, enabling it to predict at each grid of the input conditioned on meteorological variables.

To support multi-task learning, the generator shares a common backbone and branches into three task-specific output heads in the final layer. This shared-backbone design enables feature sharing across tasks while allowing each output to learn task-specific representations. The joint training encourages physically consistent predictions across precipitation and lightning fields.

To enhance the model's ability to learn the natural distribution of precipitation, we implement a discriminator, which is comprised of consecutive convolutional layers, that distinguishes between observed precipitation (the ground truth) and generated precipitation. This discriminator learns to classify labeled data as real and generated data from the generator as fake for each patch, conditioned on the input of the generator. The discriminator plays a key role in the Patch-cGAN architecture, with its loss functioning as the GAN loss that flows through the generator, encouraging it to produce sharp and realistic precipitation fields.

A schematic overview of the architecture is provided in Figure 1.

*f. Loss design*

Because the proposed architecture consists of three task-specific output head, the overall objective function is a sum of individual task losses:

$$Total\ Loss = \alpha_{precip} \cdot \mathcal{L}_{precip} + \alpha_{occ} \cdot \mathcal{L}_{occ} + \alpha_{lgt} \cdot \mathcal{L}_{lgt}, \quad (2)$$

where each loss ($\mathcal{L}$) represents loss for precipitation amount, rain probability, and lightning probability, respectively. This formulation can adopt a weighted sum version with weights ($\alpha$) to control the relative contribution of each task.

The precipitation amount loss ($\mathcal{L}_{precip}$) combines Mean Squared Error (MSE) loss with a GAN loss component, implemented as a Binary-Cross-Entropy (BCE) loss. The loss function is defined as:

$$\mathcal{L}_{precip}(G, D) = \mathbb{E}[logD(x, y)] + \mathbb{E}\left[log\left(1 - D\left(x, G(x)\right)\right)\right] + \lambda \cdot \mathcal{L}_{MSE}(G) \odot W, \quad (3)$$

where G denotes the generator mapping input variables x to output y, D denotes the discriminator. The MSE component ensures that the predicted values maintain an appropriate scale, while the GAN loss component encourages the generator to produce realistic precipitation fields that can effectively fool the discriminator. The hyperparameter $\lambda$ balances reconstruction and realism quality. Note that the MSE component is element-wisely multiplied by the weight map, $W$.

A lightning-informed weighting mechanism, which modifies the loss term to emphasize grid cells associated with convective activity, incorporates physically meaningful event awareness into the optimization process. The weight map is designed to penalize more strongly in lightning associated regions. To compensate for climatological rarity, we apply class weights proportional to the inverse square root of class frequency, which is a common approach for dealing with class imbalance (Cui et al., 2019). The weight for each class is defined as:

$$w_i = 1/\sqrt{f_i}, \tag{4}$$

where $f_i$ is climatological frequency of class, and i denotes the index of classes (here i = 0: no lightning, i = 1: lightning). The final spatial weight map $W$ is generated by mapping class weights to grid cells according to lightning occurrence labels and applying Gaussian smoothing to ensure spatial continuity and avoid abrupt transition. The sample weight map is illustrated in Figure 4.

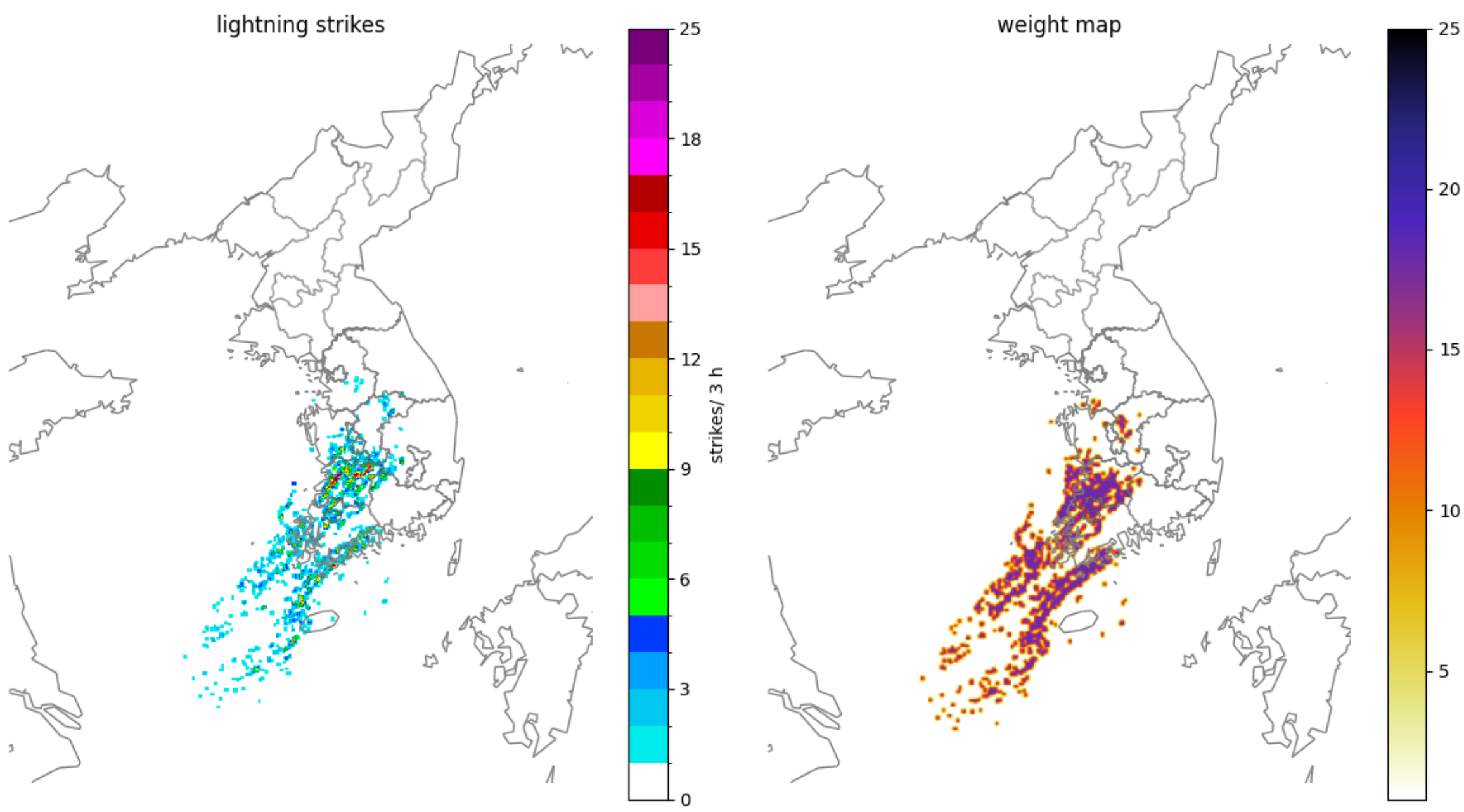

*Figure 4. An example of the lightning-informed weight map*

The rainfall occurrence ($\mathcal{L}_{occ}$) and lightning occurrence ($\mathcal{L}_{lgt}$) tasks are formulated as binary classification problems. Both adopt a Binary Cross-Entropy (BCE) loss and are also multiplied element-wisely by the lightning informed weight map, $W$.

$$\mathcal{L}_{occ} = \mathcal{L}_{BCE_occ} \odot W, \tag{5}$$

$$\mathcal{L}_{lgt} = \mathcal{L}_{BCE_lgt} \odot W, \tag{6}$$

where $\mathcal{L}_{BCE}$ indicates the BCE loss for each task.

*g. Training setup*

Training details including optimizer setting, learning rate, training epochs, and hyperparameters are described in Table 3. The loss weight for each task ($\alpha$) is equally distributed here, reflecting the physical coupling and consistency between precipitation intensity, probability, and lightning occurence. In standard Patch-cGAN architecture, the regularizer hyperparameter ($\lambda$) is typically set to 100 to prioritize pixel-wise reconstruction (Isola et al., 2017). However, because our lightning-informed weighting strategy already up-weights the MSE component in convective regions, we have set $\lambda$ to 50 (empirically determined) to maintain a balanced optimization. A larger $\lambda$ value ensures a smaller magnitude of reconstruction error, while a smaller $\lambda$ allows the adversarial loss to prioritize spatial pattern realism and details.

| | Value | Remark |
|---|---|---|
| Training epochs | 50 | - |
| Learning rate | $1e^{-5}$ | Both for Generator and Discriminator |
| Optimizer | AdamW (Loshchilov and Hutter, 2017) | Both for Generator and Discriminator<br>Weight Decay = 0.01 |
| Batch size | 32 | - |
| α | 1 | Weight for each task (precipitation, rain occurrence, and lightning probability) in Total Loss |
| $\lambda$ | 50 | Hyperparameter for Weighted MSE Loss Regularizer |
| σ | 1 | Gaussian smoothing parameter for the weight map |

*Table 3. Training setup*

## 3. Results and Verification

### *a. Verification strategies for precipitation forecasts*

A reliable Quantitative Precipitation Forecasting (QPF) model must accurately predict precipitation across all intensity categories. To assess this capability, we benchmark our proposed model against conventional NWP models and an AI-based model, GraphCast, across four precipitation intensity categories: light rain (≥ 1 mm/6 h), moderate rain (≥ 10 mm/6 h), heavy rain (≥ 20 mm/6 h), and intense rain (≥ 40 mm/6 h). Table 4 provides a summary of each model used in the verification.

For this comparison, 6-h accumulated precipitation is used to accommodate the output of GraphCast, which generates 6-h rainfall totals, whereas our proposed model is designed for a higher-frequency, shorter time-scale (3-h accumulated rain). As illustrated in Table 1, the validation set used for performance verification is entirely independent of the training set.

| | Type | Resolution | Training data | Input data | Loss Function |
|---|---|---|---|---|---|
| IFS-HRES | Global model forecast | 0.1° × 0.1° | - | - | - |
| GraphCast | AI-based model | 0.25° × 0.25° | ERA5 (1979-2019), Fine-tuned on IFS-HRES (2016-2021) | IFS-HRES | MSE Loss |
| Base-GAN | Deep-Learning based post-processing | 2km × 2km | IFS-HRES (2021-2024), Radar estimated precipitation | IFS-HRES | MSE + GAN Loss |
| EA-GAN (Event-Aware GAN) | Deep-Learning based post-processing | 2km × 2km | IFS-HRES (2021-2024), Radar estimated precipitation | IFS-HRES | MSE w/ weighted map + GAN Loss |

*Table 4. Summary of the models used for QPF benchmarking.*

To benchmark the improvement over the conventional NWP and standard AI-based forecasting, we use ECMWF's IFS-HRES as a conventional NWP baseline and also include GraphCast's precipitation forecast. GraphCast is an AI-based model that operates at a 0.25°×0.25° resolution, with initialized from IFS-HRES analysis. GraphCast, which is optimized via a multi-variable, multi-level MSE loss, is selected to represent the current performance of AI-based models trained with conventional loss formulations.

To evaluate the impact of the GAN loss and our event-aware weighting strategy, we compare two versions of the proposed architecture: the Patch-cGAN model utilizing the lightning-informed weighted loss (EA-GAN) and a baseline version utilizing no weighted loss (Base-GAN).

Before the quantitative verification, we present qualitative case studies of convective shower events and extreme precipitation cases from the Summer of 2025 in the following sections.

*b. Case studies for convective shower events*

In this section, we examine three convective shower events to demonstrate the real-world performance of our proposed loss design. We compare the Patch-cGAN model results, which are trained on high-resolution observations, with and without the event-aware loss to assess the impact of the lightning-informed weighting.

Figure 5 illustrates a thunderstorm case on 28 May 2025, driven by an upper-level cold air. Thermodynamic instability combined with afternoon adiabatic heating initiated convective storms across the eastern regions of the Korean peninsula. While the baseline model (Base-GAN) fails to detect this shower-type precipitation, our proposed loss formulation successfully enables the model (EA-GAN) to generate convective precipitation and lightning probability.

Figure 6 and 7 represent convective shower cases under the influence of a subtropical high on 21 and 22 July 2025. On both days, abundant low-level moisture advection triggered convective storms. The model without the lightning-informed loss (Base-GAN) captures only the stationary frontal precipitation over the North Korea, failing to produce the convective precipitation. In contrast, our proposed model with event-aware loss (EA-GAN) generates both the stationary frontal and the convective precipitation.

These case studies demonstrates that our proposed loss formulation effectively captures convective precipitation across varying synoptic conditions. These results suggest that GAN loss alone may not be sufficient for diagnosing rare-cases such as convective shower events.

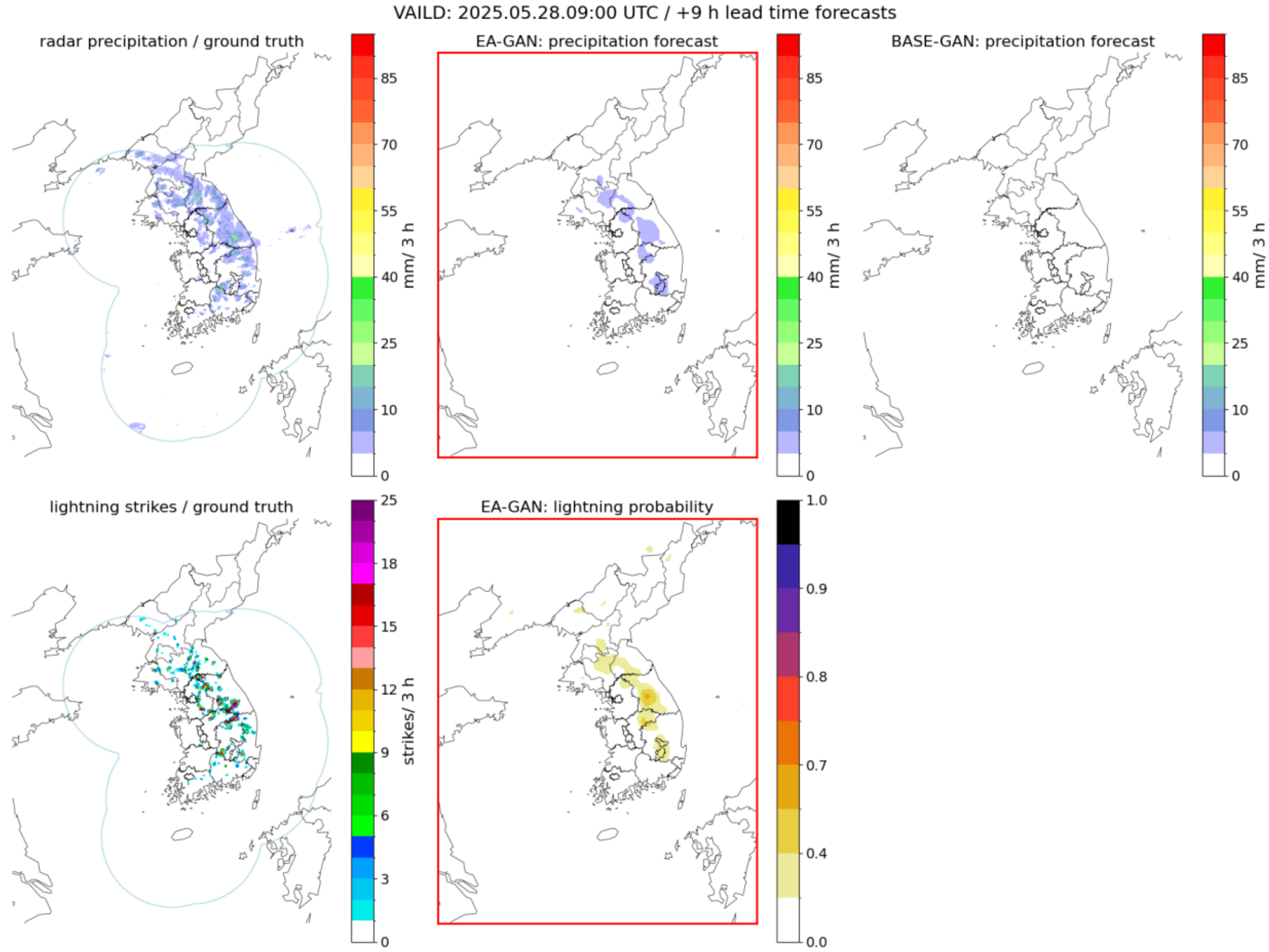

*Figure 5. Case study of a convective event at 09 UTC on 28 May 2025 (9-h lead time). (Left) Observed radar-estimated precipitation and lightning strike locations; (Middle) the proposed EA-GAN model prediction, showing 3-h accumulated precipitation and forecasted lightning probability; and precipitation forecast from the Base-GAN model (without weighted loss) (Right).*

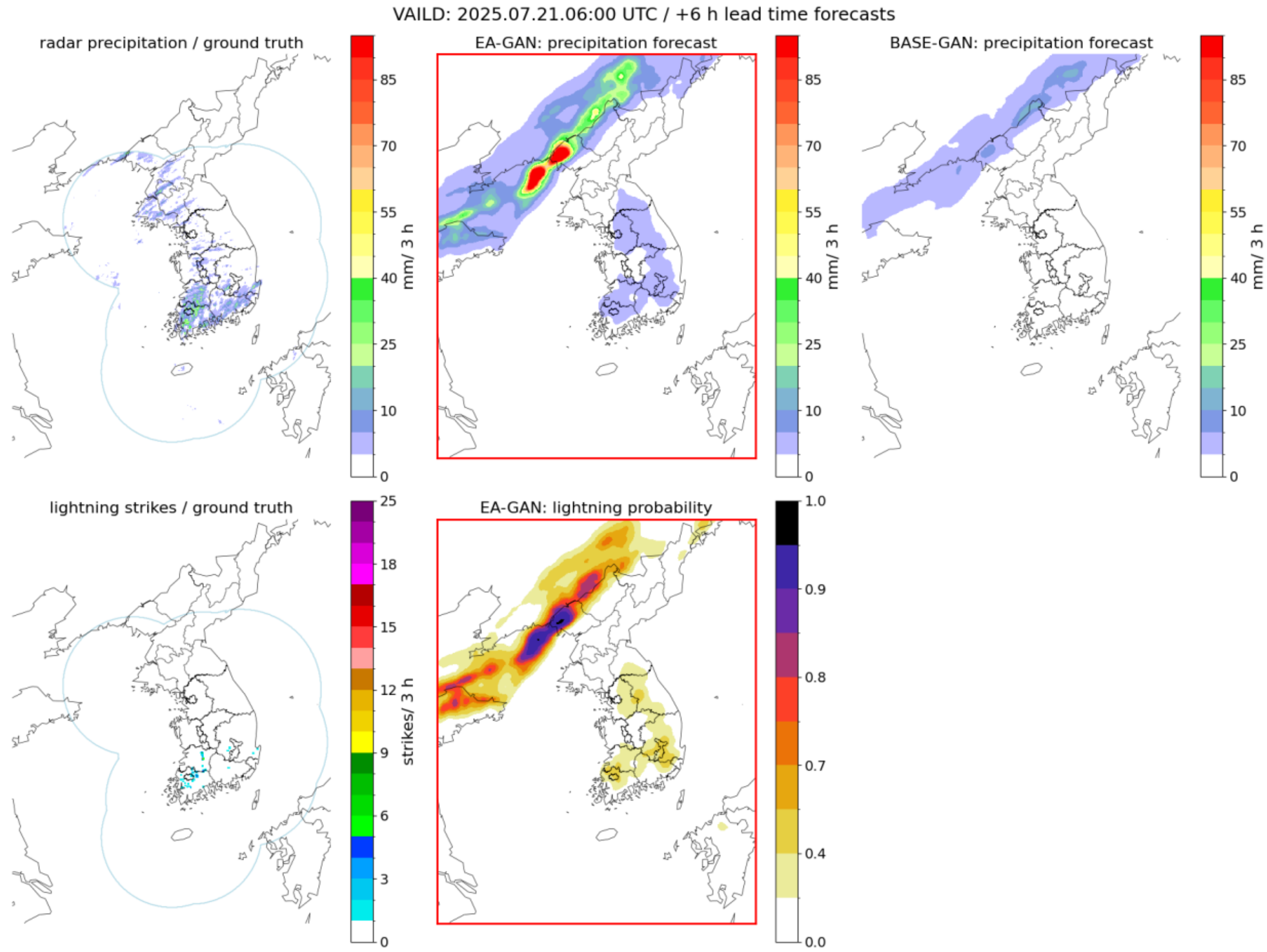

*Figure 6. Case study of a convective event at 06 UTC on 21 July 2025 (6-h lead time). (Left) Observed radar-estimated precipitation and lightning strike locations; (Middle) the proposed EA-GAN model prediction, showing 3-h accumulated precipitation and forecasted lightning probability; and precipitation forecast from the Base-GAN model (without weighted loss) (Right).*

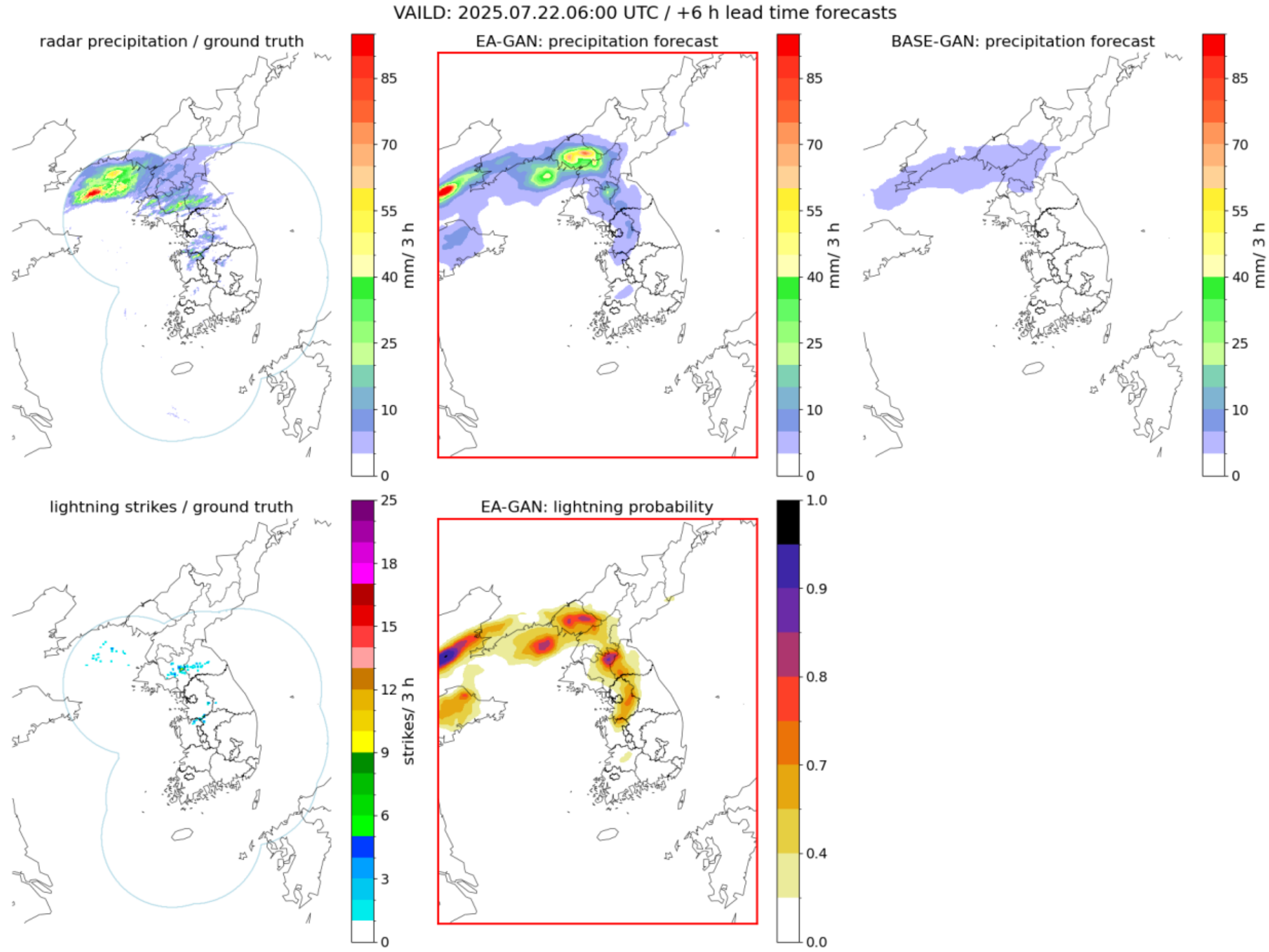

*Figure 7. Case study of a convective event at 06 UTC on 22 July 2025 (6-h lead time). (Left) Observed radar-estimated precipitation and lightning strike locations; (Middle) the proposed EA-GAN model prediction, showing 3-h accumulated precipitation and forecasted lightning probability; and precipitation forecast from the Base-GAN model (without weighted loss) (Right).*

*c. Case studies for extreme precipitation events*

Between 16 and 19 July 2025, a period of torrential rainfall affected the entire Korea peninsula. Here, we examine extreme precipitation events during this period resulted in significant property damage and societal impact.

Figure 8 illustrates a torrential rain event on 16 July 2025 in the central region of Korea, which recorded total rainfall accumulations exceeding 400 mm. A synoptic low-pressure system with abundant moisture from regions of high sea surface temperatures, triggered the development of a surface-based convective complex. Compared to the model without event-aware loss (Base-GAN), our proposed model (EA-GAN) produces higher-intensity rainfall that aligns much closely with observational data.

Figure 9 displays a convective storm event on 17 July 2025, which also resulted in over 400 mm of total rainfall in the southern region of Korea. In this case, high thermodynamic instability induced convective storms accompanying bow-echo precipitation patterns. Our loss formulation strategy enables the model (EA-GAN) to produce high-impact rainfall with lightning in the southern region of Korea.

Figure 10 represents a torrential rain event on 19 July 2025, which recorded over 350 mm of total rainfall in the southern region of Korea. The interaction between an upper-level dry air intrusion and low-level moisture regions resulted in intense rainfall. Our proposed loss-weighting strategy enables the model (EA-GAN) to generate precipitation closely matching observation.

These results demonstrates that the event-aware loss strategy can effectively predict rare, high-impact rainfall events, which GAN loss alone or conventional loss may struggle to reproduce.

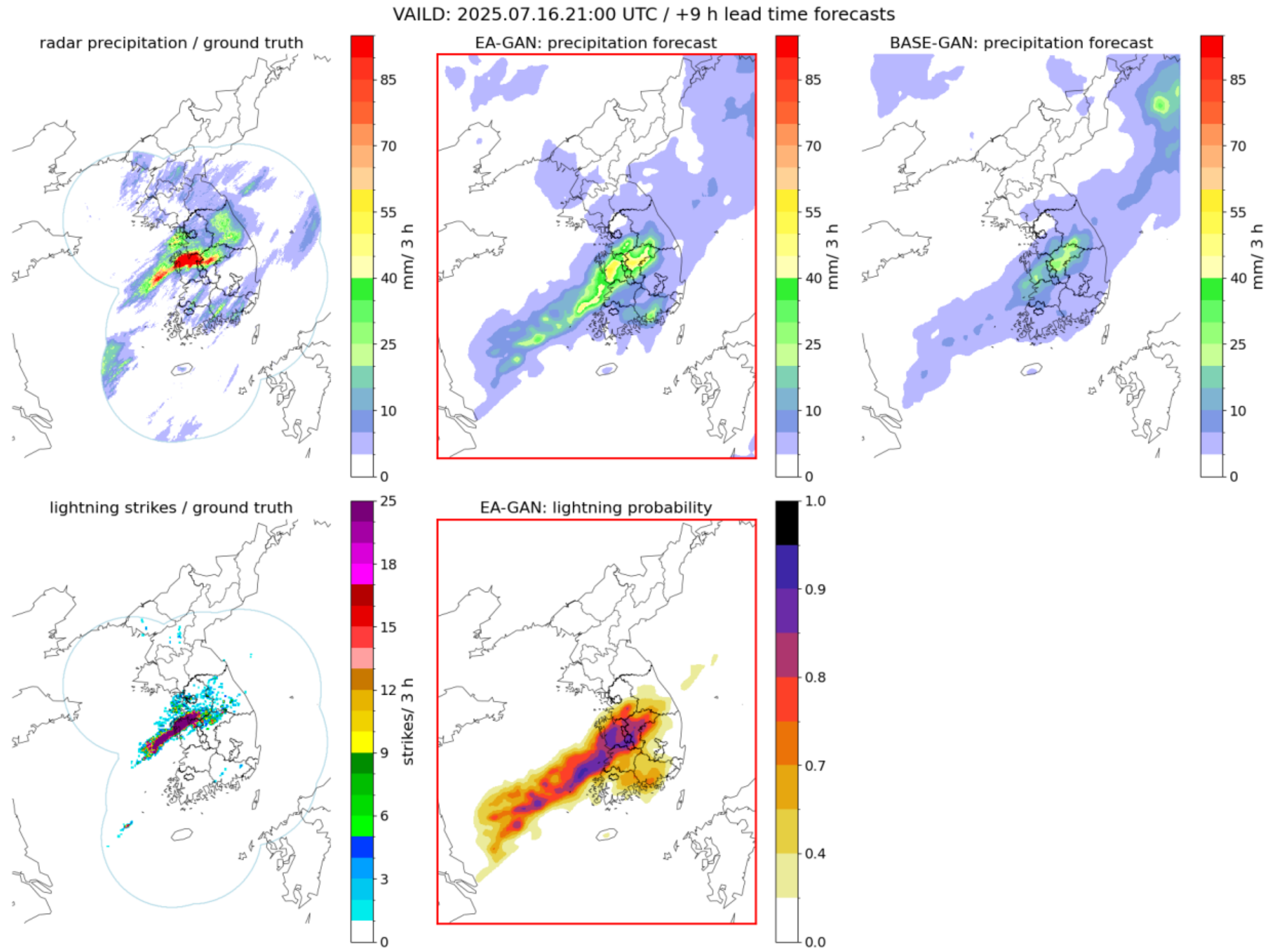

*Figure 8. Case study of an extreme event at 21 UTC on 16 July 2025 (9-h lead time). (Left) Observed radar-estimated precipitation and lightning strike locations; (Middle) the proposed EA-GAN model prediction, showing 3-h accumulated precipitation and forecasted lightning probability; and precipitation forecast from the Base-GAN model (without weighted loss) (Right).*

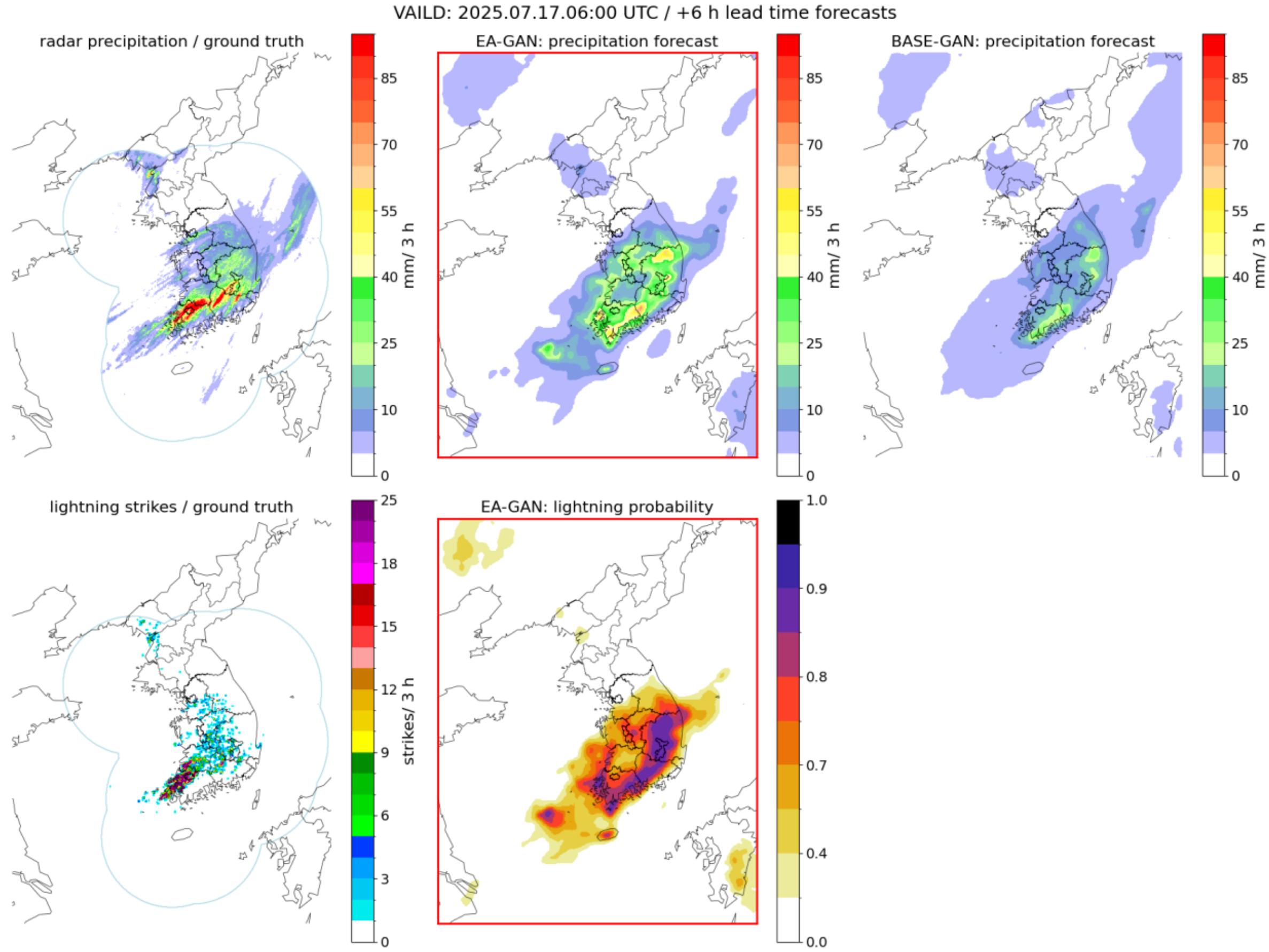

*Figure 9. Case study of an extreme event at 06 UTC on 17 July 2025 (6-h lead time). (Left) Observed radar-estimated precipitation and lightning strike locations; (Middle) the proposed EA-GAN model prediction, showing 3-h accumulated precipitation and forecasted lightning probability; and precipitation forecast from the Base-GAN model (without weighted loss) (Right).*

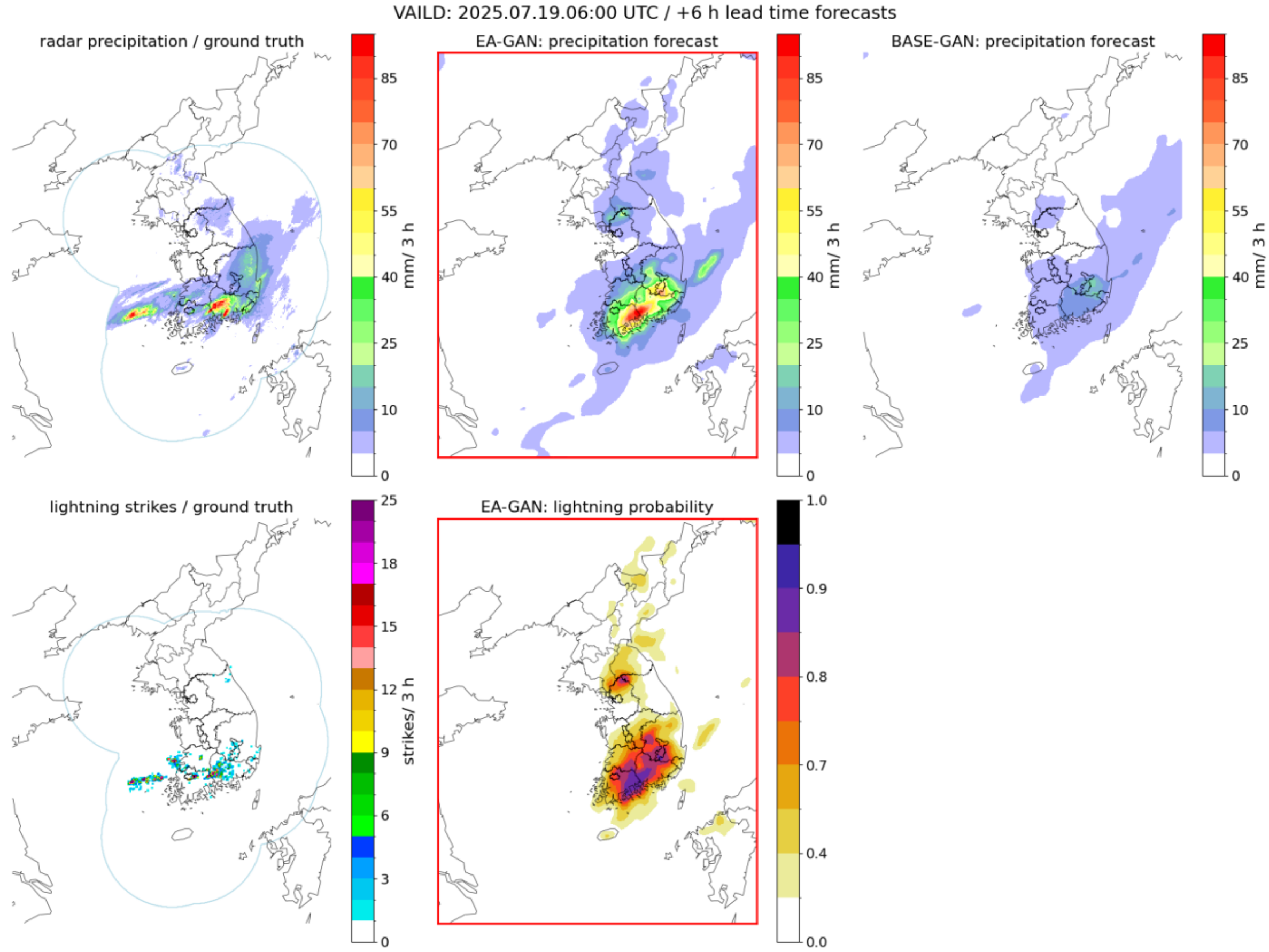

*Figure 10. Case study of an extreme event at 06 UTC on 19 July 2025 (6-h lead time). (Left) Observed radar-estimated precipitation and lightning strike locations; (Middle) the proposed EA-GAN model prediction, showing 3-h accumulated precipitation and forecasted lightning probability; and precipitation forecast from the Base-GAN model (without weighted loss) (Right).*

*d. Quantitative verification for precipitation forecasts*

We benchmark the QPF performance of various models over the summer months (June, July, and August) of 2025, using forecast lead times ranging from 6 to 120 hours from the 00 UTC runs. For a fair comparison, all models' predicted rainfall fields are re-gridded to a 2 km × 2 km resolution prior to verification. Verification is conducted using radar-estimated precipitation over the Korean Peninsula region.

Performance diagrams (Roebber, 2009) are used to benchmark the performance of each model across various metrics (Wilks, 1995), including Probability of Detection (POD, equivalent to Sensitivity; the x axis in the diagram), Success Ratio (SR, equivalent to Precision; the y axis in the diagram), Critical Success Index (CSI, equivalent to Intersection over Union; curved lines in the diagram), and Frequency Bias (FB, equivalent to Prediction Bias; the dotted diagonal lines). These diagrams allow for a comprehensive assessment, and we analyze each model's characteristics and strengths across these metrics.

In the performance diagrams (Figures 11 to 14), we aggregate the metrics for forecast days 1, 2, 3, 4, and 5 to minimize the influence of diurnal cycle effects, allowing for a more straightforward interpretation. For reference, each forecast day is marked by its corresponding number in the charts.

For the light rain threshold (1 mm/ 6 h), low-resolution models, IFS-HRES and GraphCast, tend to represent an over-forecasting bias. Specifically, the baseline AI-based model, GraphCast, shows the highest frequency bias (~1.5) among models. GAN models, Base-GAN and EA-GAN, achieve a frequency bias close to the ideal value of 1, indicating the benefits obtained from learning directly from observational data.

At the moderate rain threshold (10 mm/ 6 h), the performance of all models tends to converge. While all models maintain a frequency bias near 1, our proposed model (EA-GAN), demonstrates the best frequency bias result.

From the heavy rain threshold (20 mm/ 6 h), which is often related with convective activity, low-resolution models start to reveal their inherent weaknesses, primarily due to their low resolution and reliance on convective parameterization. Both IFS-HRES and GraphCast, show a tendency toward under-forecasting heavy rain events. While the Base-GAN model (Patch-cGAN model without weighted loss) outperforms the low-resolution models, it still exhibits an under-prediction bias at this threshold. On the other hand, EA-GAN model, which utilizes the event-aware loss formulation, achieves the highest POD and CSI, with a frequency bias near the ideal value.

For the intense rain threshold (40 mm/ 6 h), the same tendencies observed in the heavy rain category persist but with overall worse scores. At this threshold, GraphCast performs the worst among the models. This result highlights a well-known limitation of AI-based models: they tend to underestimate extreme events due to the nature of conventional MSE loss function (Rasp et al., 2023).

Figure 15 presents the significance test (Hamill, 1999) results with 95% confidence intervals, obtained from a bootstrapping with 1,000 samples. The results demonstrate that the proposed EA-GAN model achieves statistically significant improvements in CSI scores for intense rainfall (≥ 40 mm/ 6 h) compared to the baseline and conventional NWP models, particularly for lead times of 2 to 4 days. While some overlap in confidence intervals is observed at the 1-day and 5-day, the consistent upward trend in CSI underscores the robustness of the lightning-informed loss strategy.

The verification results indicate that the EA-GAN model, through its weighted-loss strategy, effectively generates precipitation fields that closely match observations. By maintaining a frequency bias near 1 across all categories, with particular improvements in the heavy rain and intense rainfall categories often associated with deep convection, the proposed framework successfully overcomes the under-forecasting limitation inherent in traditional NWP and standard AI models with conventional loss. This outcome highlights the advantage of our model learning precipitation features directly from observational data with the proper loss strategy.

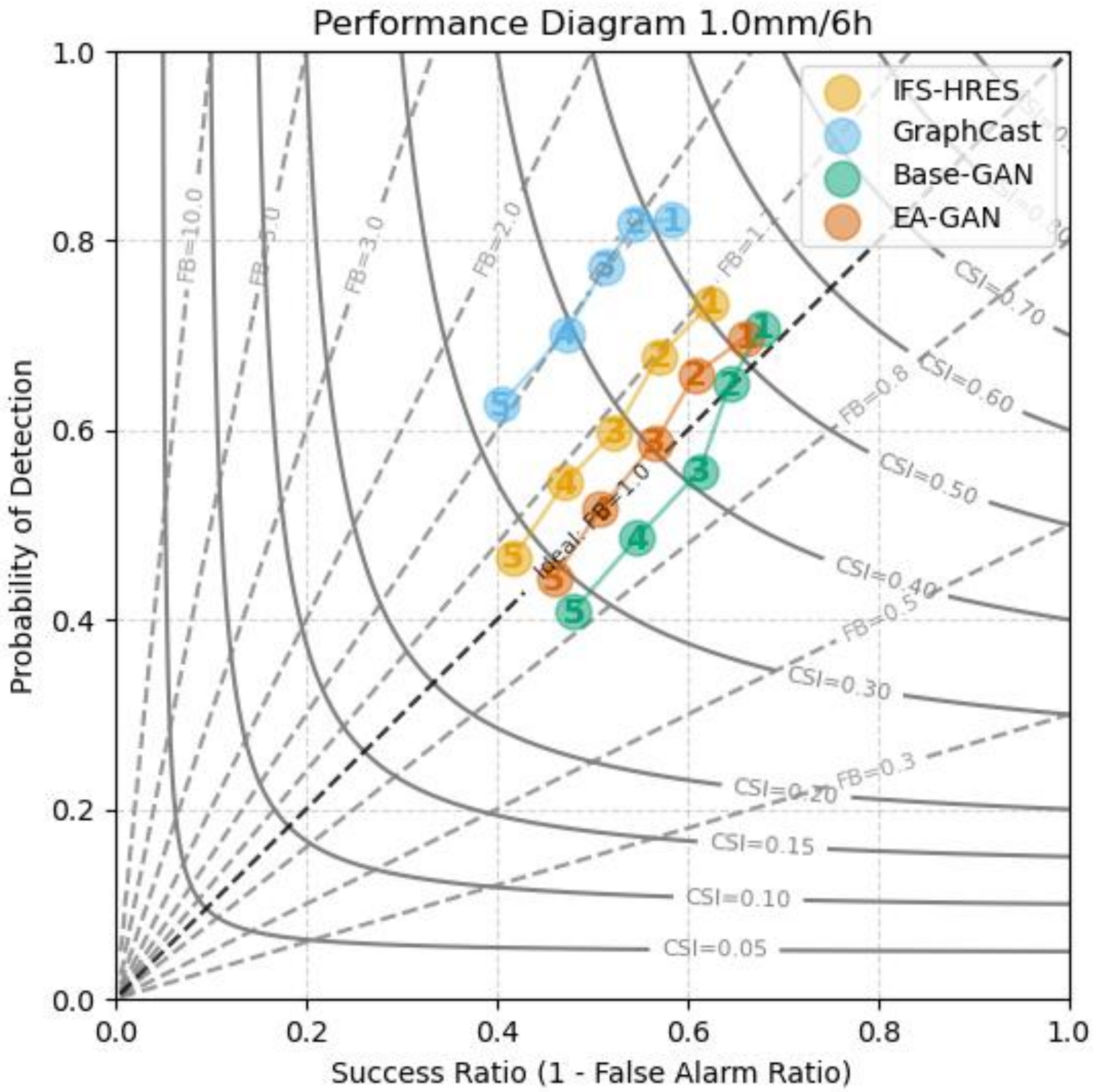

*Figure 11. Performance diagram for each model at the 1 mm/6 h rainfall threshold, from day 1 to day 5 forecasts during the summer of 2025. Each forecast day is indicated by its corresponding number in the chart.*

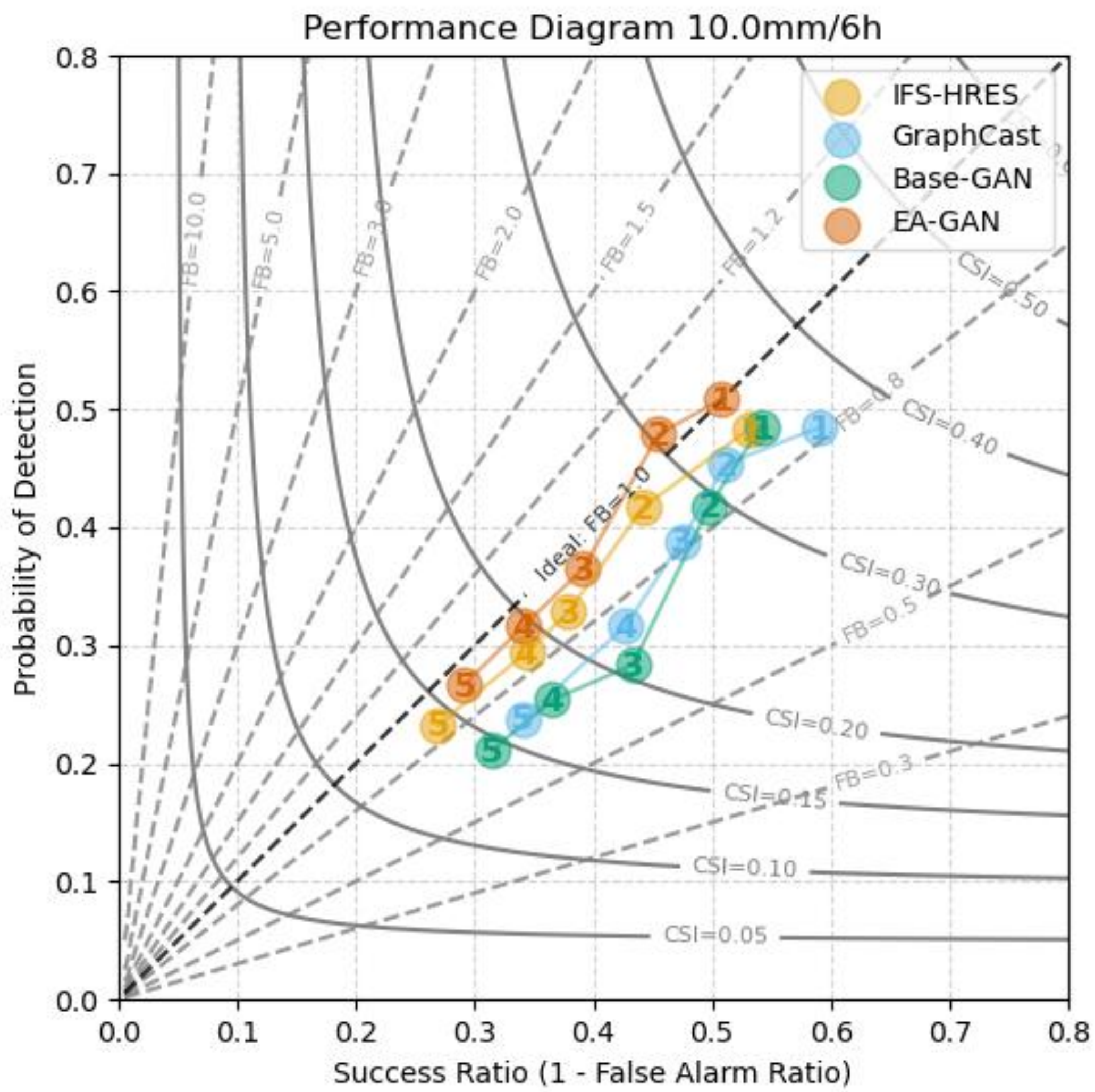

*Figure 12. Performance diagram for each model at the 10 mm/6 h rainfall threshold, from day 1 to day 5 forecasts during the summer of 2025. Each forecast day is indicated by its corresponding number in the chart.*

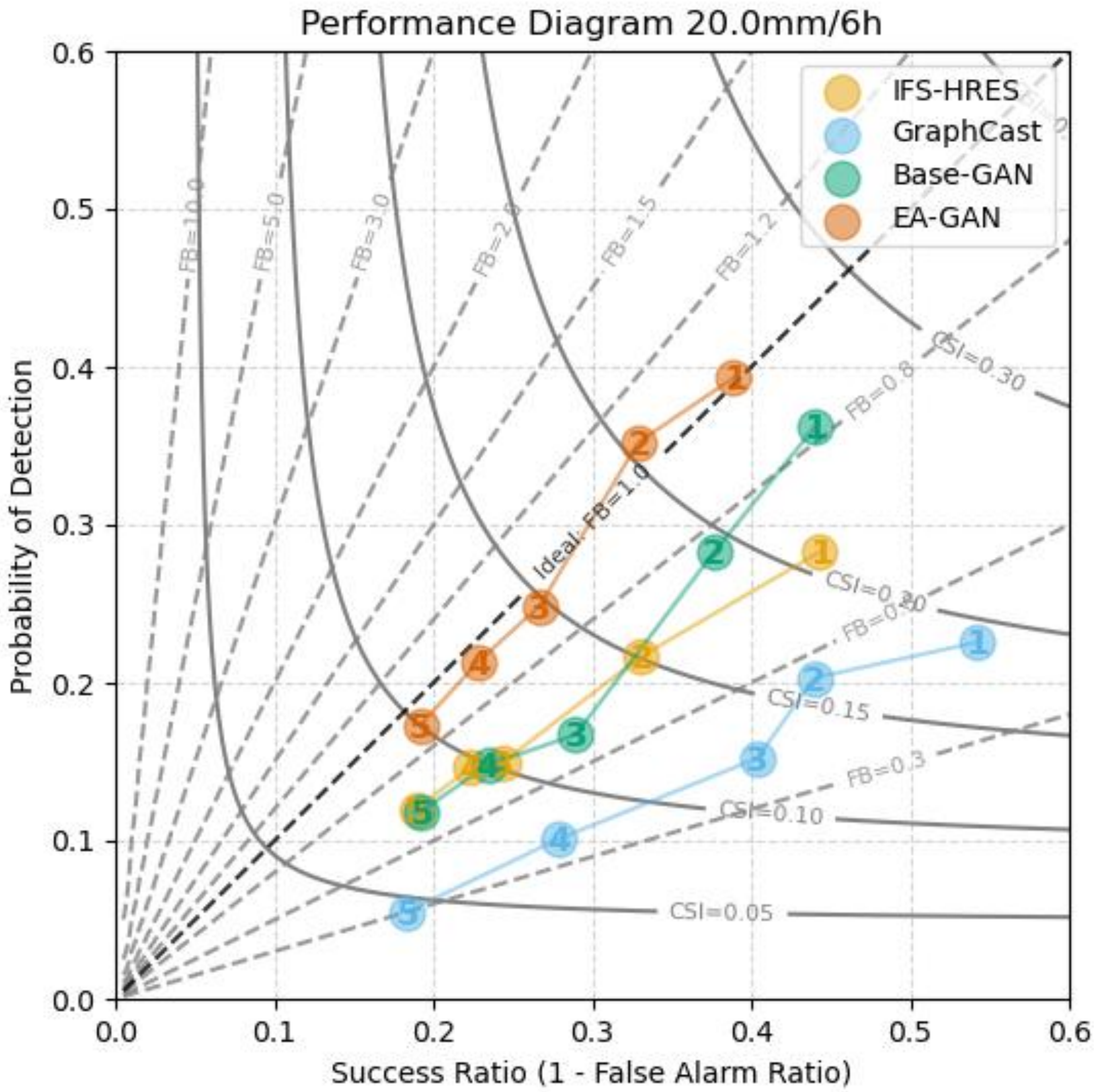

*Figure 13. Performance diagram for each model at the 20 mm/6 h rainfall threshold, from day 1 to day 5 forecasts during the summer of 2025. Each forecast day is indicated by its corresponding number in the chart.*

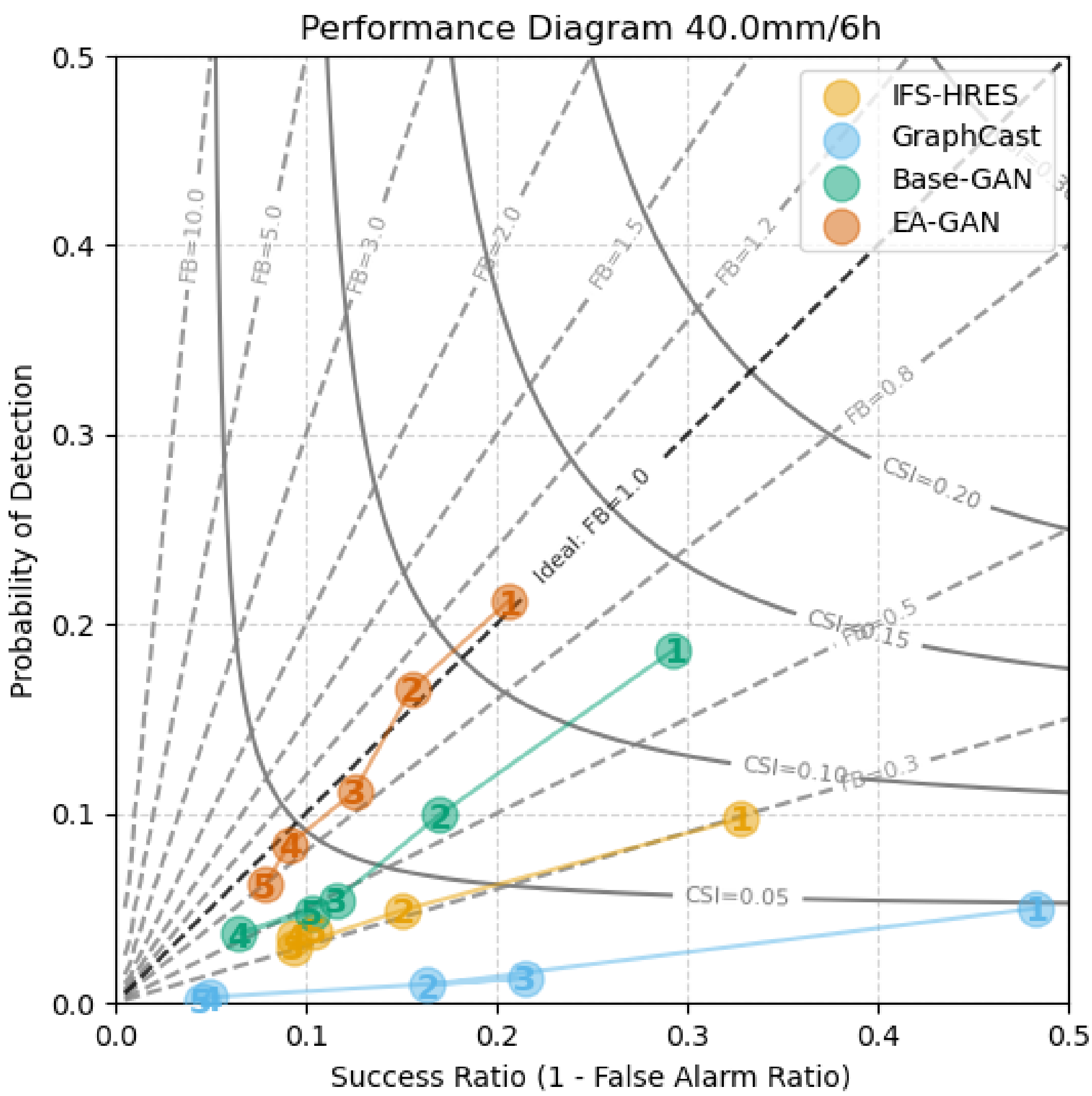

*Figure 14. Performance diagram for each model at the 40 mm/6 h rainfall threshold, from day 1 to day 5 forecasts during the summer of 2025. Each forecast day is indicated by its corresponding number in the chart.*

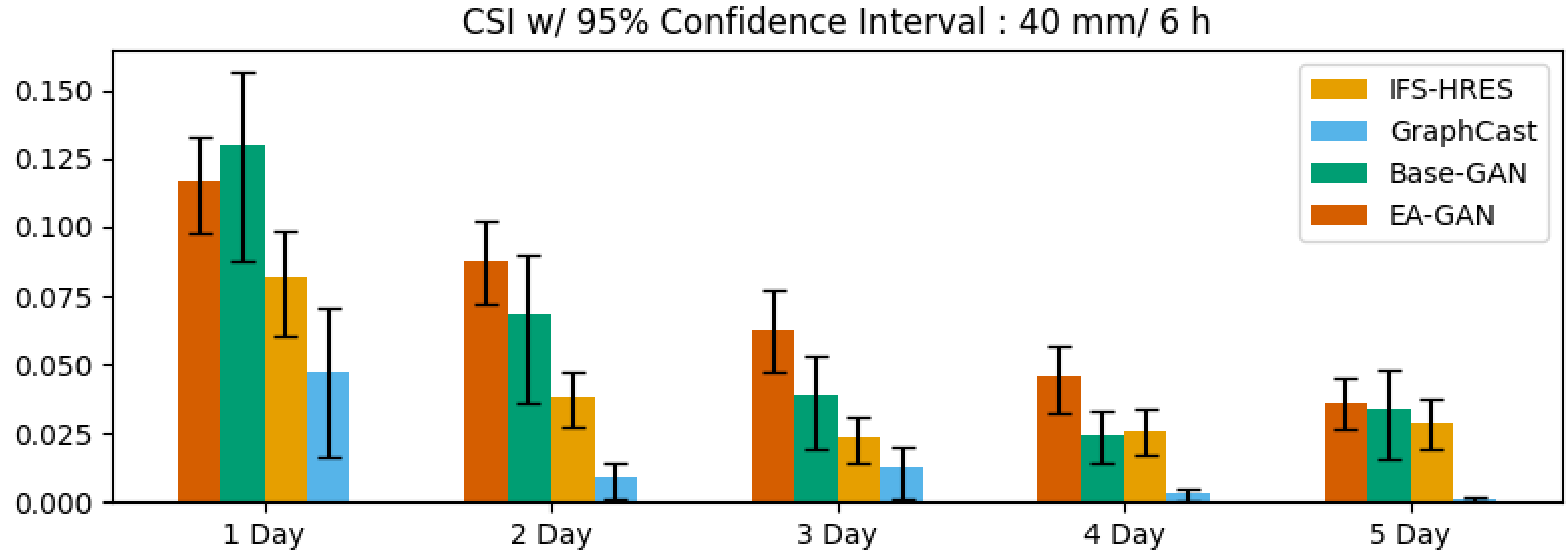

*Figure 15. CSI results for each model at the 40 mm/6 h rainfall threshold with 95% confidence interval using bootstrapping (1,000 samples), from day 1 to day 5 forecasts during the summer of 2025.*

*e. Verification for lightning forecasts*

To benchmark lightning prediction performance, we compare a lightning parameterization and an instability-index-based guidance against our data-driven machine learning outcomes. The first baseline, Total Lightning Flash Density (IFS-LFD), is a parameterization that estimates lightning using convective hydrometeor amounts, Convective Available Potential Energy (CAPE), and convective cloud base height derived from the IFS-HRES forecast (Lopez, 2016).

The second baseline, the Cloud Physics Thunder Parameter (CPTP), referred to here as IFS-CPTP, is a diagnostic index used to identify thunderstorm potential. It is calculated using vertical sounding information, including CAPE, the lifting condensation level, and the equilibrium level temperature (Bright et al., 2004). Conventionally, regions where CPTP exceeds 1 (unitless) are identified as areas with thunderstorm potential.

Although the GAN architecture is not designed for lightning detection, our proposed EA-GAN model is trained to identify potential lightning regions as a classification task. To ensure a fair comparison with baseline methods, we select a 75% probability threshold for lightning prediction. At this threshold, the model achieves a frequency bias near the ideal value of 1, ensuring a balanced prediction that avoids over-prediction or under-prediction of lightning.

|  | Type | Resolution | Threshold | Remark |
|---|---|---|---|---|

| IFS-LFD (Lightning Flash Density) | Lightning parametrization | 0.1° × 0.1° | $1\ /100\ km^2/$ h | Total Lightning Flash Densities based on convective scheme |
|---|---|---|---|---|
| IFS-CPTP (Cloud Physics Thunder Parameter) | CAPE based index | 0.1° × 0.1° | 1 (unitless) | Sounding information based guidance |
| EA-GAN | Deep-Learning model output | 2km × 2km | 75% | Data-driven machine-learning-based guidance |

*Table 5. Summary of the models used for lightning prediction benchmarking.*

Figure 16 presents a performance diagram for lightning forecasts during the Summer of 2025. While both the IFS-LFD and EA-GAN results exhibit balanced predictions with a frequency bias near 1, IFS-CPTP over-predicts lightning occurrence nearly three times more often than observations and yielding the lowest SR. This discrepancy indicates that vertical sounding information alone is insufficient for classifying lightning events, as unstable thermodynamic conditions do not always materialize into deep convection.

Our EA-GAN model demonstrates clear superiority over existing guidance, maintaining both the highest POD and SR for lead times up to 4 days. This outcome underscores the advantage of a data-driven approach for capturing such rare and complex phenomena. While the verification results for IFS-LFD are modest in terms of detection, this guidance provides a comparative advantage in lightning intensity information.

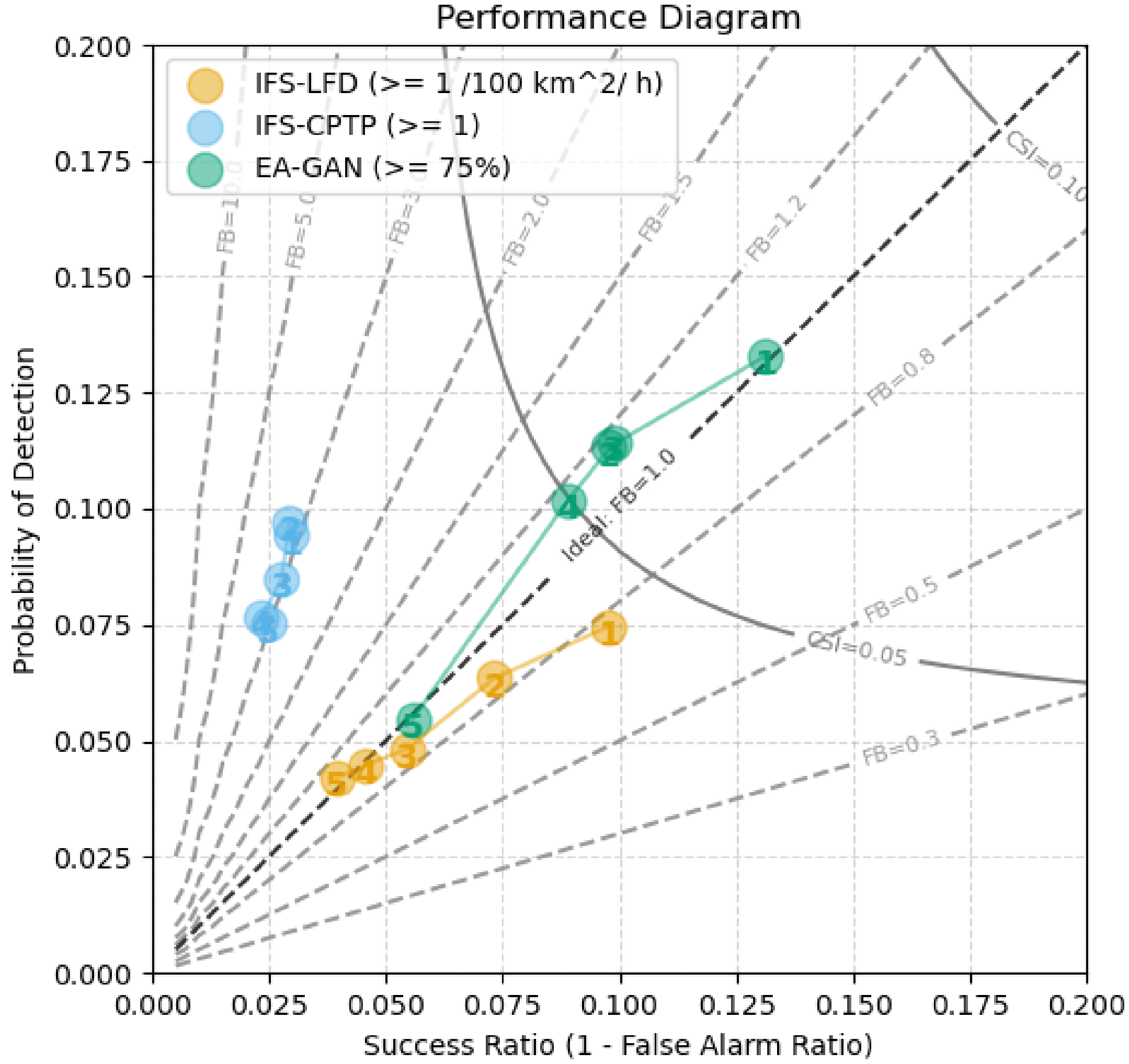

*Figure 16. Performance diagram for lightning forecasts, from day 1 to day 5 forecasts during the summer of 2025. Each forecast day is indicated by its corresponding number in the chart.*

## 4. Summary and Discussion

Based on the benchmark results and case studies, we demonstrate that the proposed EA-GAN model effectively generates convective events while maintaining a frequency bias near the ideal value of 1. This lightning-informed loss strategy, which explicitly enables the model to focus on convective signals during the training stage, delivers measurable improvements in both the detection of convective showers and the quantitative precipitation forecasting of extreme events compared to conventional loss formulations.

The baseline GAN model (Base-GAN), which is trained on high-resolution radar observations, struggles to detect localized convective shower events. This difficulty likely stems from the inherent ambiguity of localized features, where similar meteorological conditions may yield contradictory observational outcomes (i.e., rain or no rain). However, by applying a weight map that more heavily penalizes errors where lightning occurred, the optimization process is guided to favor the generation of precipitation in convective environments.

These findings suggest that the lightning-informed loss strategy successfully guides the model to learn the specific meteorological conditions associated with convective activity. In this context, the approach can be interpreted as a data-driven parametrization schemes for convective precipitation; however, unlike traditional NWP schemes, it is learned directly from the real-world observations. This event-aware loss formulation serves as a remedy for the limitations inherent in conventional loss functions, which often exhibit a persistent under-prediction bias for extreme events and struggle with localized convective shower detection when trained on high-resolution data.

Our proposed EA-GAN model also demonstrates the improvement in lightning prediction over a lightning parametrization method and instability-index-based methods, such as CPTP. By leveraging the advantage of learning directly from the lightning observations, our model achieves both a higher POD and a higher SR than existing guidance. These results underscore the superiority of data-driven methods over physical parametrization schemes for predicting rare and complex phenomena such as lightning.

While this study is conducted over the Korea peninsula region, the framework is designed for scalability. This method can be extended to further regions where lightning observations are available, including areas covered by Geostationary Lightning Mappers (Goodman et al., 2013) satellite, offering a pathway toward global convective forecasting.

*Acknowledgments*

This work was conducted in collaboration with the Korea Meteorological Administration (KMA) and the Distributed Intelligence and Systems Laboratory at Chungnam National University, Korea. The authors express their gratitude to the Korea Meteorological

Administration for providing the high-performance GPU computing resources essential for conducting the experiments and model training in this study.

*Data Availability Statement.*

The code used in this study is available at https://github.com/hunter3789/Deep-Learning-QPF. The training and validation datasets used in this publication are accessible at https://osf.io/ehwmv/files/osfstorage.